\documentclass[fleqn,10pt]{wlscirep}
\usepackage[utf8]{inputenc}
\usepackage[T1]{fontenc}
\usepackage{lineno}

\newcommand{\corr}[1]{\textcolor{black}{#1}}

\title{Detection and tracking of barchan dunes using Artificial Intelligence\\
\normalsize{\textcolor{blue}{Accepted Manuscript version of: Cúñez, E.A., Franklin, E.M., Detection and tracking of barchan dunes using Artificial Intelligence. Scientific Reports, 14, 18381, 2024. https://doi.org/10.1038/s41598-024-67893-y. This article is licensed under a Creative Commons Attribution 4.0 International License:  https://creativecommons.org/licenses/by/4.0/}}}

\author[1]{Esteban A. C\'u\~nez}
\author[1,*]{Erick M. Franklin}
\affil[1]{Faculdade de Engenharia Mec\^anica, Universidade Estadual de Campinas (UNICAMP), Campinas-SP, 13083-860, Brazil}
\affil[*]{erick.franklin@unicamp.com}

\begin{abstract}
Barchans are crescent-shape dunes ubiquitous on Earth and other celestial bodies, which are organized in barchan fields where they interact with each other. Over the last decades, satellite images have been largely employed to detect barchans on Earth and on the surface of Mars, with AI (Artificial Intelligence) becoming an important tool for monitoring those bedforms. However, automatic detection reported in previous works is limited to isolated dunes and does not identify successfully groups of interacting barchans. In this paper, we inquire into the automatic detection and tracking of barchans by carrying out experiments and exploring the acquired images using AI. After training a neural network with images from controlled experiments where complex interactions took place between dunes, we did the same for satellite images from Earth and Mars. We show, for the first time, that a neural network trained properly can identify and track barchans interacting with each other in different environments, using different image types (contrasts, colors, points of view, resolutions, etc.), with confidence scores (accuracy) above 70\%. Our results represent a step further for automatically monitoring barchans, with important applications for human activities on Earth, Mars and other celestial bodies.
\end{abstract}

\begin{document}

\flushbottom
\maketitle
%
%
\thispagestyle{empty}

\section*{Introduction}

Barchans are dunes of crescent shape with horns pointing downstream, which are formed mainly under one-directional flows and when the amount of available sand is limited \cite{Bagnold_1}. These bedforms are frequently found on Earth (both in aquatic and eolian environments) and Mars, having in common the same morphology, but presenting different scales \cite{Hersen_1}: they are much larger and slower on Mars, where the scales are up to one kilometer for the length and millenniums for the turn-over time (although their average length has recently been found \cite{Rubanenko2} to be of the order of 200 m, and the turn-over time on the north pole to be of the order of a century \cite{Chojnacki2}), than under water, whose scales are tens of centimeters and minutes \cite{Claudin_Andreotti}. On terrestrial deserts, the scales \corr{are up to hundreds of meters and years}. However, it is not always easy to clearly identify barchans and measure their dimensions based on images, since they are organized in barchan fields in which they migrate over long distances while interacting with each other \cite{Hersen_2, Hersen_5, Kocurek, Genois, Genois2, Assis, Assis2, Assis3}. In addition, the fluid flow often presents seasonal variations, affecting the morphology of dunes \cite{Parteli5, Courrech2, Gadal}. Therefore, it is common to observe barchans that touch each other (colliding dunes), that are highly asymmetric, and that shed small barchans, for instance. Despite these difficulties, barchan dunes are still a bedform of much interest since they can be used to deduce information about the atmosphere of planets and moons based on satellite images, such as the existence of an atmosphere that is or has been capable of mobilizing sediments (otherwise there would not exist barchans), the mean direction of winds, and even the flow strength (from stability analyses) \cite{Claudin_Andreotti}. In the particular case of Martian barchans, these inferences represent in some cases mean winds that have blown over the last millenniums (given the turn-over time of large Martian barchans), something that the satellites and sensors in locus cannot measure.

Over the last decades, with better resolution satellites being launched and orbiting Earth and also Mars, dunes on both planets have been monitored with reasonable accuracy \cite{Bourke2, Silvestro, Fenton}. In particular, the detection of barchans on Earth and Mars based on satellite images have been largely employed. The first works detecting barchans used non-machine learning detection \cite{Tsoar, Bourke2, Tsoar2, Zhang_2}, meaning that the dunes were identified, classified and measured (morphology and, sometimes, displacement) with an algorithm specially written for these purposes, instead of being trained for identification; however, with the advance of Machine and Deep Learnings (ML and DL, respectively), automatic detection based on computer training has become more common \cite{Azzaoui, Carrera}. As pointed out by Rubanenko et al. \cite{Rubanenko}, those works, based on Support Vector Machine \cite{kowalczyk_sup} or R-Vine classifiers, were accurate for automatically detecting and classifying, but not for segmenting and outlining individual barchans.

Recently, Rubanenko et al. \cite{Rubanenko} made use of Mask R-CNN (Regional Convolutional Neural Network) \cite{He}, which detects objects while simultaneously generating a segmentation mask, for automatically detecting, classifying and outlining barchan dunes on Mars and Earth. To minimize false detection of barchans and derive the trends for wind and sand transport, they focused their training on isolated barchan dunes, which they carried out for 1076 images and surpassed 70\% of accuracy (mean average precision $mAP$, see section Methods). Afterward, they applied the Mask R-CNN to 137111 images of the Martian surface extracted from a CTX global mosaic from the Mars Reconnaissance Orbiter (MRO) Context Camera (CTX) dataset. With those images, they mapped large regions of the Martian surface and found that around 60\% of dune fields on the northern hemisphere are covered with barchans, while only 30\% of fields on the southern hemisphere are covered with barchans. Finally, they applied the same training to satellite images from Earth and obtained reasonable accuracy (which can be improved by inverting the image colors or performing new training). Later, Rubanenko et al. \cite{Rubanenko2} explored the barchans identified and outlined in Ref. \cite{Rubanenko} for probing the assumption that m-scale ripples found on Mars are the result of a hydrodynamic instability. Their measurements showed that the lengths of both small barchans and m-size ripples decrease with increasing the atmosphere density, following, in addition, a power-law predicted by a hydrodynamic analysis, which corroborates the initial assumption.

Although recent works, in especial Rubanenko et al. \cite{Rubanenko, Rubanenko2}, increased the accuracy in detecting and outlining individual barchans in satellite images, variations in the crescentic shape (due to barchan-barchan interactions and seasonal winds) and the existence of other types of dunes that are also curved (parabolic dunes, for instance) hinder the automatic detection and outline in many cases. In particular, to the best of the authors' knowledge, current trained networks do not detect successfully groups of interacting barchans, mainly when they are touching or superposing each other (barchan-barchan collisions). In addition, because previous works were conducted for single images, it remains to be proved that CNNs (Convolutional Neural Networks) can track the detected barchans and update their outline along a sequence of frames (or movie). The automatic detection of barchans undergoing complex interactions (such as barchan-barchan collisions) is relevant for many reasons. One of them is that it opens the possibility for updating the number of barchans on the surface of planets (Mars, for example), which were possibly underestimated in previous works (since they computed isolated barchans only), and determining their location, orientation and concentration \cite{Rubanenko}. This information can be useful for estimating the direction and strength of local winds, determining the regimes of sand transport and accumulation, and estimating the effects of global changes on Earth based on the dynamics of dunes \cite{Baas}. Other reason is the possibility for predicting the future of barchan fields based on barchan-barchan interaction maps, such as those from Assis and Franklin \cite{Assis} (or, in the same way, deducing the ancient past of such fields), or the interaction of dunes with dune-size obstacles based on the corresponding maps, such as shown in \corr{Assis et al.} \cite{Assis4}: by training a CNN with interacting patterns measured in laboratory, the trained CNN might predict the same kind of interaction in the field based on satellite images. Here again, this information can be used for estimating desertification as an effect of climate change \cite{Baas}, and predicting if constructions are under imminent threat of being overtaken by sand \cite{WoodTV}. Another application would be the yearly monitoring of dune motion for estimating the sand cover on Earth and testing climate models.
	
In this paper, we inquire into the automatic detection, classification, outline, and tracking of barchans in different environments by carrying out experiments and exploring with DL the images of both individual and groups of barchans. We made use of the existing python library YOLO (You Only Look Once) for training a neural network with images from experiments where complex interactions took place between dunes, and, afterward, did the same for satellite images from Earth and Mars. We show, for the first time, that the trained network can identify, classify, outline, and track dunes interacting with each other in different environments, using different image types (contrasts, colors, points of view, resolutions, etc.), with confidence scores (estimated accuracy for each detected object) within 70 and 90\% and mean average precision that reaches 99\%. The trained CNN opens new possibilities for updating the number of barchans on the surface of planets (by considering also those undergoing complex interactions) and predicting the future of barchan fields (based on barchan-barchan interaction maps \cite{Assis} and satellite images). Our results represent a step further for automatically monitoring barchans and understanding their dynamics, with important applications for human activities, such as mitigating disasters on Earth and exploring Mars.

\section*{Results}

\begin{figure}[ht]
	\centering
	\includegraphics[width=.99\linewidth]{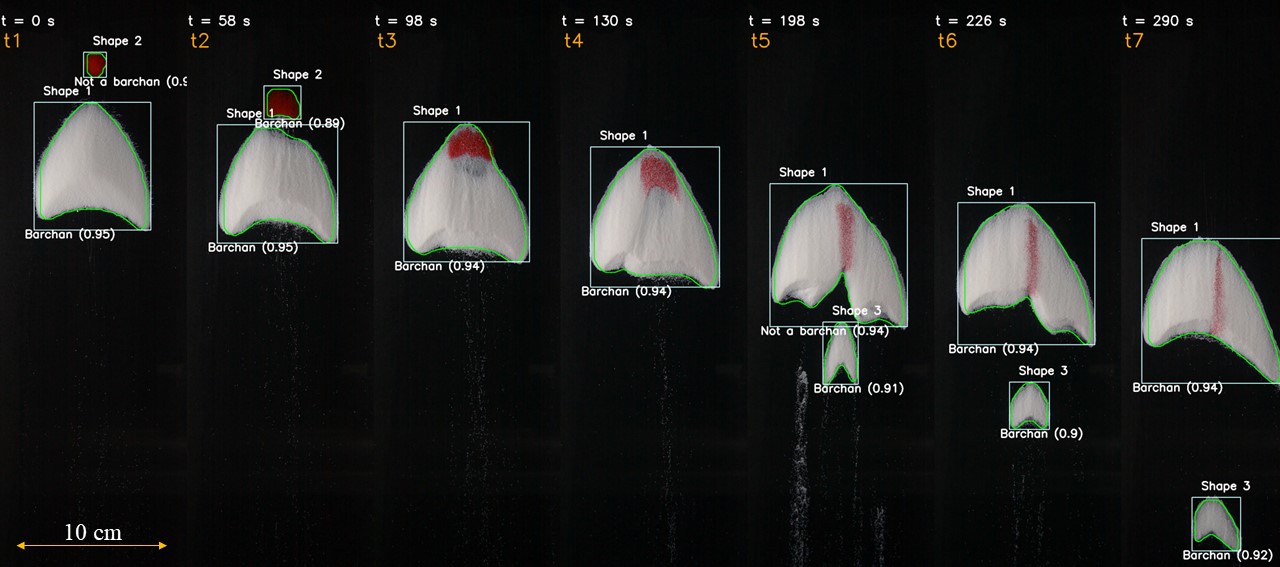}
	\caption{Snapshots placed side by side of two barchans interacting in a pattern called \textit{exchange} \cite{Assis}. The images were taken from the open repository \cite{Supplemental_binary_grl} created by Assis and Franklin \cite{Assis}, from which we selected some time instants shown on the top of each snapshot. The instants were numbered from $t1$ to $t7$ (shown in orange on the top), a length scale is shown on the bottom left, the boxes enclosing the identified objects are shown in white, and the outline of objects are shown in green. Each object has a label (Shape 1 to Shape 3) which is kept until the last image, and the classes \textit{Barchan} and \textit{Not a barchan} are shown for each object with the corresponding confidence score. In the images, water flow is from top to bottom and the grains forming one of the bedforms (that which was initially upstream) are red in order to track them along images \cite{Assis}.}
	\label{fig:detection}
\end{figure}

After training the network with a certain number of images from the experiments (see section Methods for more details), we applied the trained network to identify, classify, outline and track dunes that interacted between each other in other experiments. One advantage of training the CNN with images from experiments with subaqueous barchans is that they contain the whole interaction sequence, allowing the accurate identification of bedforms during labeling. Besides, because in the experiments the flow conditions, grains' properties, and terrain conditions are well controlled, we can ascertain which is the exact barchan-barchan interaction that is going on. We applied the trained object detection, for instance, to some of the experiments of Assis and Franklin \cite{Assis} (open dataset available \cite{Supplemental_binary_grl}), in which the initially upstream and downstream dunes consisted of grains of different colors. For example, Fig. \ref{fig:detection} shows snapshots placed side by side of two barchans as the smaller barchan (red dune, initially upstream) collides with the larger one (white, downstream). They merge and, after some time has elapsed, eject a small barchan whose size is similar to the barchan initially upstream, but consisting of different grains (this pattern is know as \textit{exchange} \cite{Assis}). Time instants are shown on the top, and they are numbered from $t1$ to $t7$ (also shown on the top). We observe that the trained CNN can detect the objects (in this case, bedforms), which are bounded by white boxes in the figure. Each object has an label assigned (Shape 1 to Shape 3), is classified as \textit{Barchan} or \textit{Not a barchan} with an accuracy (confidence score) higher than 0.9 (see the section Methods for a description of the computation of the accuracy shown in the images), and the outline is determined (in green lines). The assigned label is kept constant for each object along the images, so that after Shape 2 is absorbed by Shape 1 it disappears, and the new ejected barchan is identified by another label (Shape 3). The identification, classification, outline and tracking work well independent of the color of objects, evincing the ability of the trained CNN in following each object along frames. Interestingly, the trained CNN is able to correctly identify as one single barchan the initially merged dune ($t$ = 98 to 130 s), even if the red grains continue with a barchan-like shape. This is a major result of this training, since object detection codes usually have problems in correctly identifying this kind of interaction \cite{Assis, Assis2, Assis3}. This opens new possibilities for computing more accurately the number of barchans appearing in satellite images (since barchans undergoing complex interactions can now be detected and outlined), and also for predicting the future configurations of barchan fields \cite{Assis}.

\begin{figure}[ht]
	\centering
	\includegraphics[width=.8\linewidth]{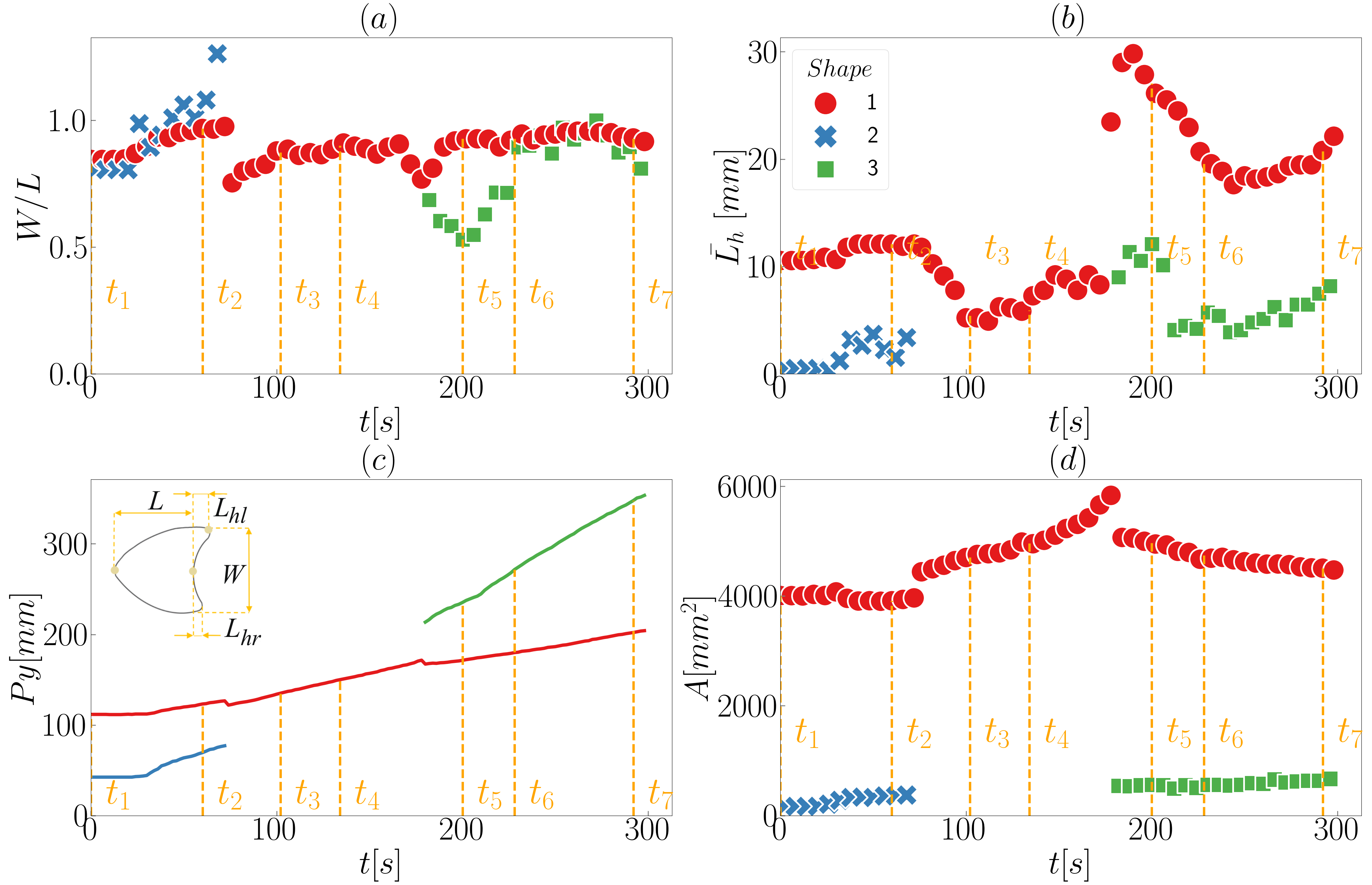}
	\caption{Time evolution of morphology and position of interacting bedforms for the case shown in Fig. \ref{fig:detection}: (a) aspect ratio $W/L$; (b) mean horn length $\overline{L}_h$; (c) longitudinal position $P_y$ of the centroid of bedforms; and (d) Area $A$ of the horizontal projection of the bedform surface. Time instants corresponding to the snapshots of Fig. \ref{fig:detection} are shown in all panels. The insert in panel (c) shows the dimensions of barchan dunes considered in this figure: their length $L$, width $W$, length of the right horn $L_{hr}$, and that of the left horn $L_{hl}$. The mean horn length is $\overline{L}_h$ $=$ $(L_{hr} + L_{hl})/2$.}
	\label{fig:morphodynamics}
\end{figure}

\begin{figure}[ht]
	\centering
	\includegraphics[width=.4\linewidth]{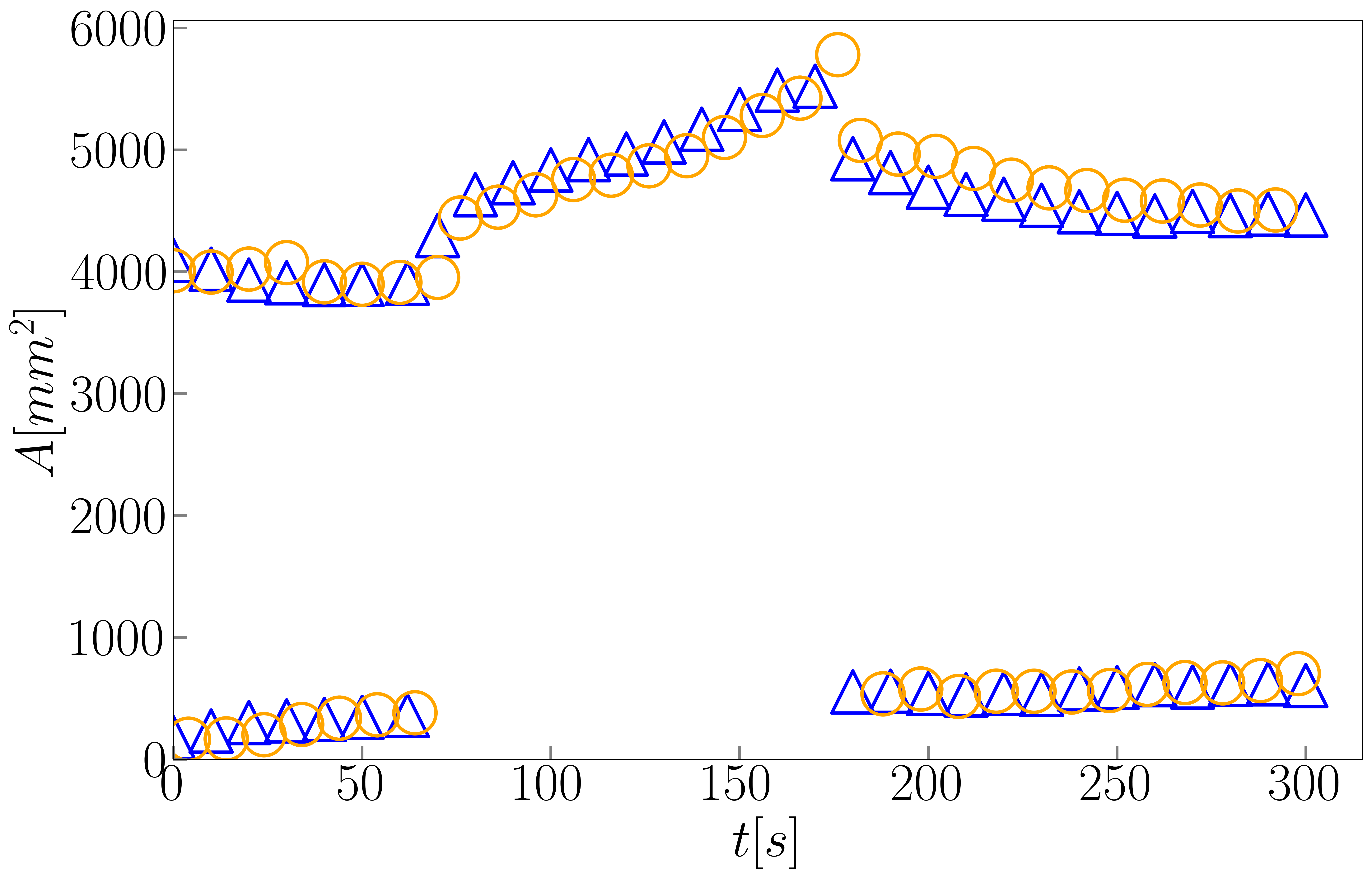}
	\caption{Comparison between automatic and manual detections: time evolution of the area $A$ (horizontal projection) of bedforms for the case shown in Fig. \ref{fig:detection} (orange circles) and the results computed manually by Assis and Franklin \cite{Assis} (blue triangles).}
	\label{fig:area_comparison}
\end{figure}

In order to evaluate if positions and outlines are accurately detected and tracked, we computed the length $L$, width $W$, horns' length $L_h$ and surface area $A$ of barchans. The lengths, width and area were computed as defined in Assis and Franklin \cite{Assis, Assis2}, and the mean horn length $\overline{L}_h$ was computed as the average of both horns. We note that $A$ is the area bounded by the outlines generated by trained CNN, corresponding thus to the surface area of the dune projected in the horizontal plane. For the exchange case of Fig. \ref{fig:detection}, Figs. \ref{fig:morphodynamics}a--d show the \corr{time evolution of} the aspect ratio $W/L$, mean horn length $\overline{L}_h$, longitudinal position $P_y$, and projected area $A$, and the time instants corresponding to those of snapshots of Fig. \ref{fig:detection} are shown in the panels. We observe that these quantities are in good agreement with the results shown in Assis and Franklin \cite{Assis}, in which a conventional (non-machine learning) detection code was used. In particular, Fig. \ref{fig:morphodynamics}d can be compared directly with Fig. 3f of Ref. \cite{Assis}, which we show in Fig. \ref{fig:area_comparison}, and the agreement is perfect (deviations of less than 5\%). We applied the same trained CNN to many other experiments, in especial the other cases reported in Assis and Franklin \cite{Assis} (images and movies of which are available in an open repository \cite{Supplemental_binary_grl}), and the results were as good as those of Figs. \ref{fig:detection} and \ref{fig:morphodynamics} (the results for other experiments are available in the Supplementary Information, as well as movies showing the tracking of individual dunes along frames). Obtaining results in good agreement with dedicated (non-machine learning) codes implies that the latter are no longer necessary for future experiments with subaqueous dunes, and that AI (Artificial Intelligence) can be successfully used for processing image sequences taken from drones or satellites.

\begin{figure}[ht]
	\centering
	\includegraphics[width=.99\linewidth]{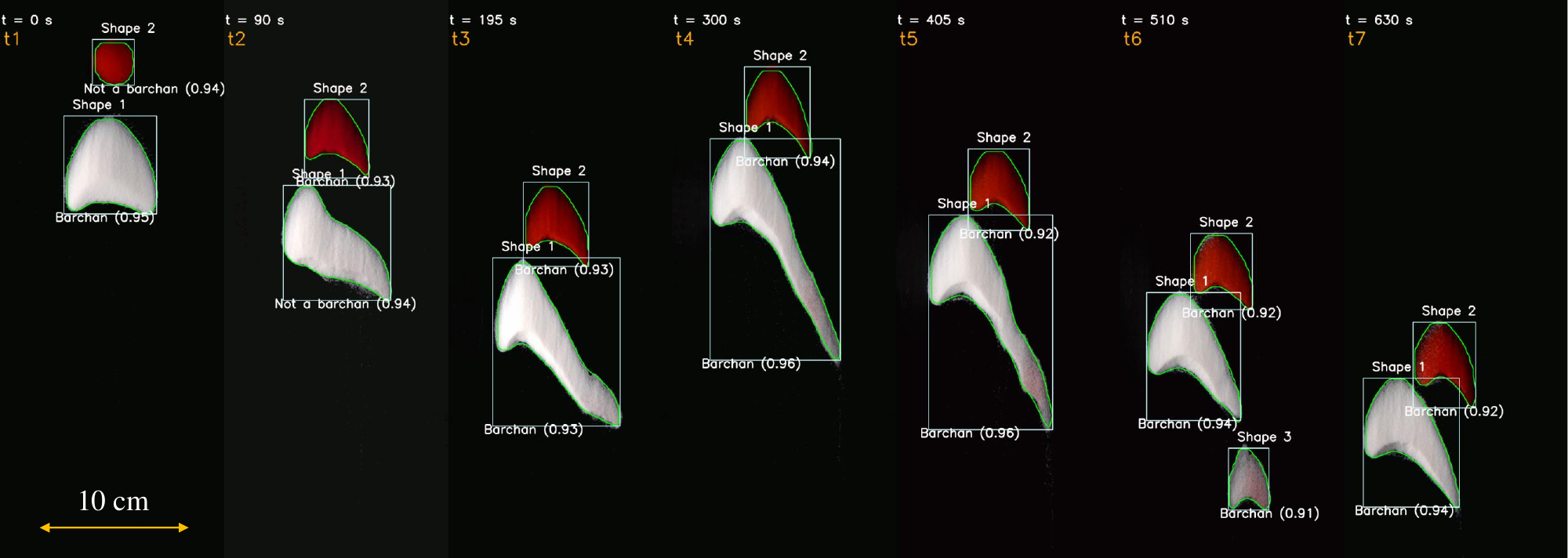}
	\caption{Snapshots placed side by side of two barchans interacting in a pattern called \textit{fragmentation} \cite{Assis}. The images were taken from the open repository \cite{Supplemental_binary_grl} created by Assis and Franklin \cite{Assis}, from which we selected some time instants shown on the top of each snapshot. The symbols and labels are as in Fig. \ref{fig:detection}. In the images, the water flow is from top to bottom and the grains forming one of the bedforms (that which was initially upstream) are red in order to track them along images \cite{Assis}.}
	\label{fig:detection2}
\end{figure}

It is worth noting that in some experiments the camera was displaced (it was mounted on a traveling system) to maintain the barchans in its field of view. In those cases, even with the spatial reference changing abruptly between two images, the trained CNN tracked correctly each barchan, as can be seen in some of the movies of the Supplementary Information. This can be also seen in Fig. \ref{fig:detection2}, in which the camera was displaced between $t$ = 195 and 300 s (in the tests, it was displaced at some point between those instants in an abrupt way, i.e., the test stopped, the camera was displaced, and the test re-started). Figure \ref{fig:detection2} shows snapshots side by side of two barchans that interact with each other, and, at some point in time, one of them ejects a small barchan (the \textit{fragmentation} pattern described in Assis and Franklin \cite{Assis}). The dunes are correctly outlined and tracked, even if with the camera displacement the barchans are (wrongly) seen in the frame as in upstream positions with respect to previous frames, showing the CNN robustness. The morphological parameters obtained in this case and others are available in the Supplementary Information, and are in agreement with Ref. \cite{Assis}.

\begin{figure}[h!]
	\centering
	\includegraphics[width=.65\linewidth]{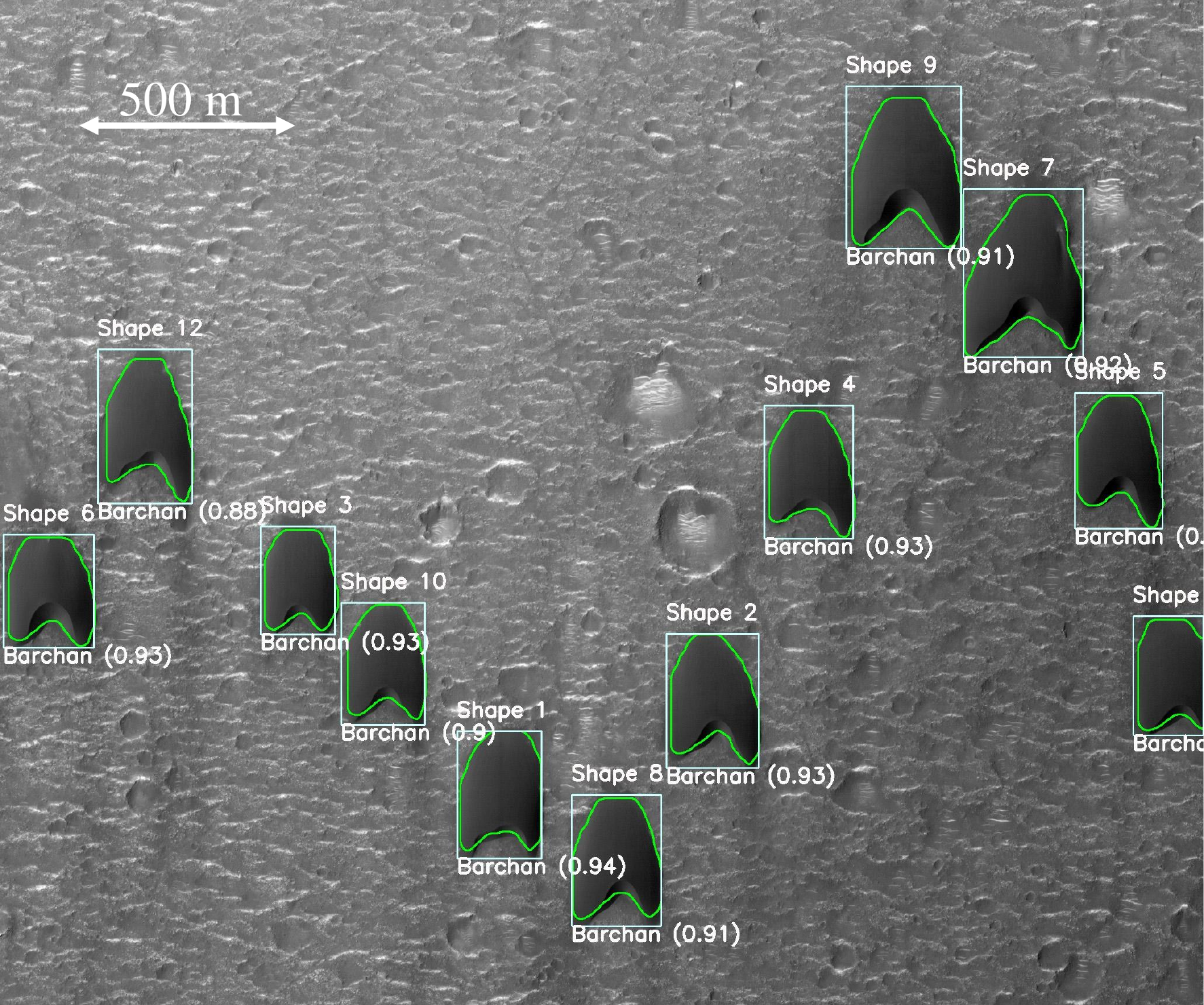}
	\caption{HiRISE image \cite{Supplemental_HIRISE} showing a field of individual barchans on the surface of Mars: 23.190$^\circ$ latitude (centered), 339.585$^\circ$ longitude (East), spacecraft altitude 287.3 km. Courtesy NASA/JPL-Caltech/UArizona. The detection boxes, class type, confidence score, and outline are superposed with the image.}
	\label{fig:satellite}
\end{figure}

\begin{figure}[h!]
	\centering
	\includegraphics[width=.65\linewidth]{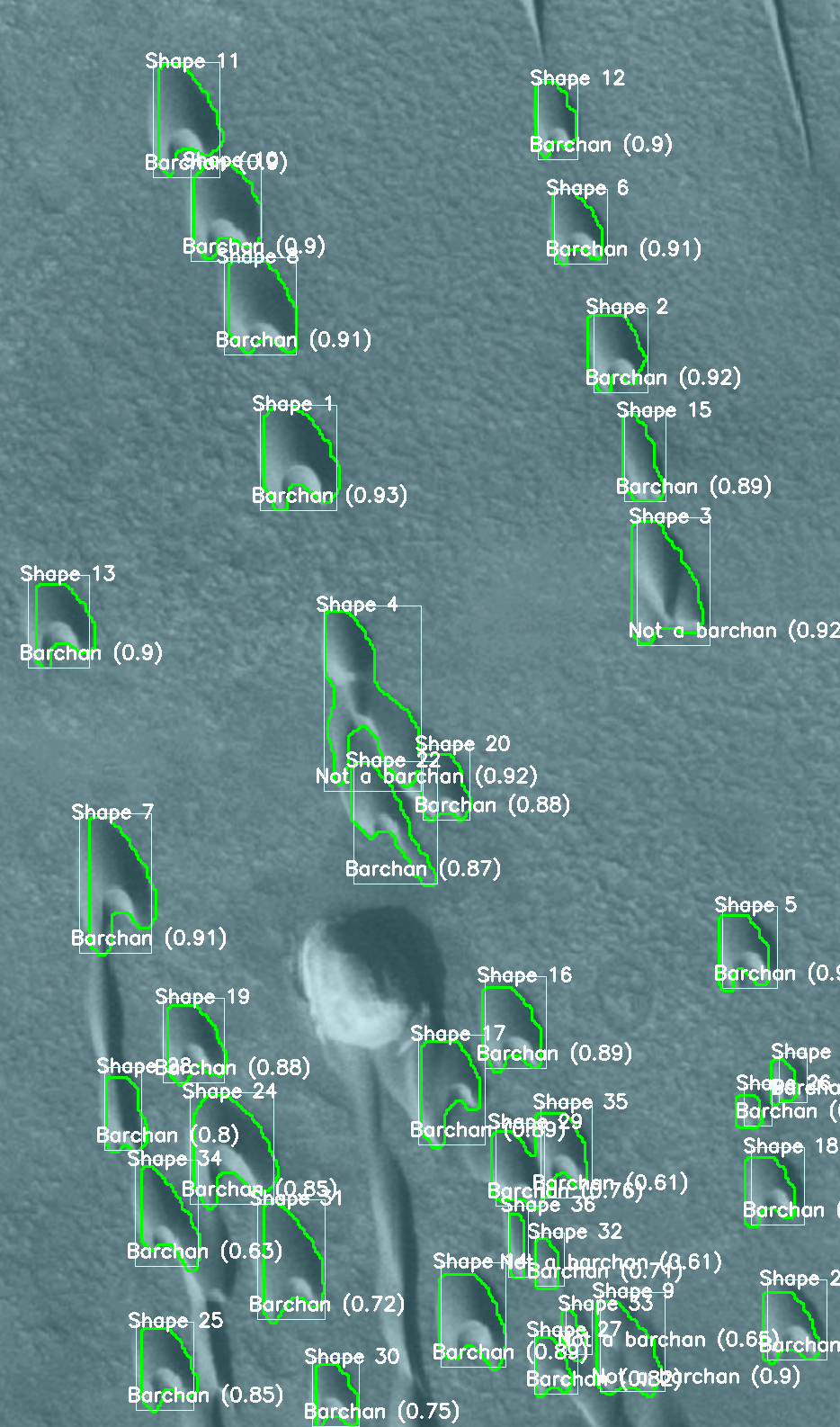}
	\caption{Medium-resolution image (4 m per px) from CTX \cite{Supplemental_CTX} showing a field of barchans undergoing complex interactions on the surface of Mars: -41.488$^\circ$ latitude (centered), 44.589$^\circ$ longitude (East), spacecraft altitude 253.8 km. Courtesy NASA/JPL/MSSS/The Murray Lab. The detection boxes, class type, confidence score, and outline are superposed with the image. The detections using the HiRISE image are available in the Supplementary Information.}
	\label{fig:satellite2}
\end{figure}

\begin{figure}[h!]
	\centering
	\includegraphics[width=.65\linewidth]{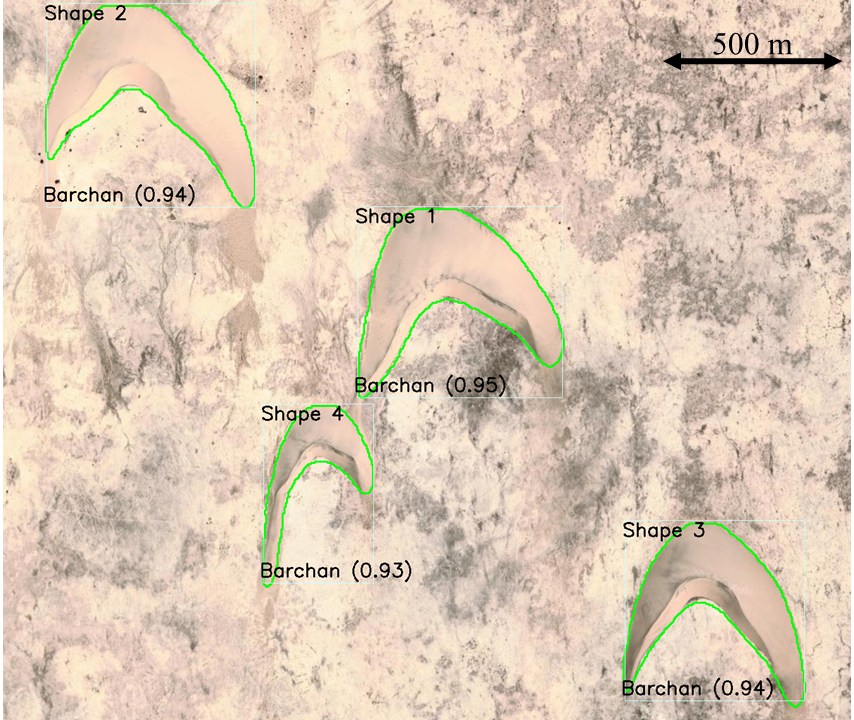}
	\caption{Satellite image showing a barchan field on Earth. This images corresponds to 24.836$^\circ$ latitude (centered),  51.311$^\circ$ longitude (East), on Qatar, with spacecraft altitude of 28.0 m. Courtesy Google Earth. The detection boxes, class type, confidence score, and outline are superposed with the image.}
	\label{fig:satellite3}
\end{figure}

\begin{table}[!ht]
	\begin{center}
		\caption{Morphological properties of dunes detected, classified, labeled, and outlined in Fig. \ref{fig:satellite}: Label (shape number), class type, length $L$, width $W$, mean horn length $\overline{L}_h$, projected area $A$, and longitudinal and transverse components of the centroid position, $P_x$ and $P_y$, respectively.}
		\label{tab:satellite1}
		\begin{tabular}{c c c c c c c c}			
			\hline		
			Shape & Class & $L$ & $W$ & $\overline{L}_h$ & $A$ & $P_x$ & $P_y$ \\
			$\ldots$    &  $\ldots$    & m &  m &  m & m$^2$ & m & m \\ \hline
			1 & Barchan & 226 & 183 & 41 & 40179 & 1130 & 1782 \\
			2 & Barchan & 217 & 200 & 64 & 42256 & 1607 & 1574 \\
			3 & Barchan & 188 & 166 & 38 & 30441 & 674 & 1307 \\
			4 & Barchan & 221 & 195 & 47 & 39556 & 1825 & 1063 \\
			5 & Barchan & 218 & 196 & 54 & 40181 & 2522 & 1030 \\
			6 & Barchan & 188 & 196 & 51 & 37143 & 112 & 1330 \\
			7 & Barchan & 275 & 267 & 81 & 65984 & 2308 & 625 \\
			8 & Barchan & 213 & 200 & 57 & 41305 & 1387 & 1936 \\
			9 & Barchan & 250 & 250 & 84 & 61080 & 2041 & 392 \\
			10 & Barchan & 209 & 178 & 44 & 35578 & 868 & 1492\\
			11 & Barchan & 213 & 150 & 35 & 32703 & 2637 & 1518 \\
			12 & Barchan & 238 & 195 & 61 & 43157 & 333 & 961 \\  \hline\hline
		\end{tabular}
	\end{center}
\end{table}

\begin{table}[!ht]
	\begin{center}
		\caption{Morphological properties of dunes detected, classified, labeled, and outlined in Fig. \ref{fig:satellite2}: Label (shape number), class type, length $L$, width $W$, mean horn length $\overline{L}_h$, projected area $A$, and longitudinal and transverse components of the centroid position, $P_x$ and $P_y$, respectively.}
		\label{tab:satellite2}
		\begin{tabular}{c c c c c c c c}			
			\hline		
			Shape & Class & $L$ & $W$ & $\overline{L}_h$ & $A$ & $P_x$ & $P_y$ \\
			$\ldots$    &  $\ldots$    & m &  m &  m & m$^2$ & m & m \\ \hline			
			 1 & Barchan & 194 & 187 & 53 & 33850 & 725 & 1116 \\
			2 & Barchan & 145 & 145 & 31 & 19948 & 1500 & 857 \\
			3 & Not a barchan & 262 & 176 & 24 & 38059 & 1616 & 1431 \\
			4 & Not a barchan & 308 & 217 & 116 & 51840 & 1033 & 1796 \\
			5 & Barchan & 130 & 123 & 47 & 17485 & 1812 & 2325 \\
			6 & Barchan & 148 & 123 & 30 & 16430 & 1403 & 553 \\
			7 & Barchan & 233 & 165 & 70 & 36051 & 282 & 2160 \\
			8 & Barchan & 189 & 167 & 33 & 30297 & 634 & 751 \\
			9 & Not a barchan & 211 & 161 & 19 & 27392 & 1526 & 3308 \\
			10 & Barchan & 183 & 167 & 36 & 29604 & 548 & 509 \\
			11 & Barchan & 207 & 161 & 41 & 28309 & 449 & 284 \\
			12 & Barchan & 145 & 101 & 28 & 13923 & 1352 & 286 \\
			13 & Barchan & 145 & 145 & 41 & 20814 & 148 & 1522 \\
			14 & Barchan & 174 & 165 & 46 & 28321 & 1147 & 3228 \\
			15 & Barchan & 196 & 101 & 4 & 15383 & 1564 & 1130 \\
			16 & Barchan & 167 & 145 & 34 & 23161 & 1244 & 2517 \\
			17 & Barchan & 178 & 145 & 41 & 25586 & 1088 & 2656 \\
			18 & Barchan & 126 & 123 & 33 & 15705 & 1876 & 2909 \\
			19 & Barchan & 145 & 145 & 40 & 20572 & 469 & 2552 \\
			20 & Barchan & 183 & 106 & 10 & 16324 & 1204 & 2048 \\
			21 & Barchan & 141 & 139 & 28 & 17977 & 1925 & 3246 \\
			22 & Barchan & 167 & 209 & 75 & 23915 & 943 & 1997 \\
			23 & Barchan & 101 & 66 & 6 & 5732 & 1914 & 2647 \\
			24 & Barchan & 189 & 209 & 68 & 37545 & 559 & 2810 \\
			25 & Barchan & 150 & 139 & 36 & 19820 & 401 & 3345 \\
			26 & Barchan & 79 & 66 & 1 & 4600 & 1830 & 2720 \\
			27 & Barchan & 132 & 95 & 17 & 10556 & 1346 & 3345 \\  
			28 & Barchan & 167 & 101 & 15 & 14883 & 302 & 2731 \\
			29 & Barchan & 195 & 111 & 17 & 16417 & 1273 & 3069 \\
			30 & Barchan & 128 & 104 & 32 & 12558 & 817 & 3416 \\
			31 & Barchan & 233 & 150 & 42 & 30404 & 705 & 3090 \\
			32 & Barchan & 123 & 57 & 1 & 6203 & 1335 & 3094 \\
			33 & Not a barchan & 88 & 59 & 17 & 4963 & 1401 & 3268 \\
			34 & Barchan & 192 & 145 & 40 & 21825 & 407 & 2984 \\
			35 & Barchan & 188 & 99 & 18 & 17515 & 1413 & 3035 \\
			36 & Not a barchan & 150 & 37 & 3 & 5288 & 1262 & 3048 \\	
			\hline\hline
		\end{tabular}
	\end{center}
\end{table}

\begin{table}[!ht]
	\begin{center}
		\caption{Morphological properties of dunes detected, classified, labeled, and outlined in Fig. \ref{fig:satellite3}: Label (shape number), class type, length $L$, width $W$, mean horn length $\overline{L}_h$, projected area $A$, and longitudinal and transverse components of the centroid position, $P_x$ and $P_y$, respectively.}
		\label{tab:satellite3}
		\begin{tabular}{c c c c c c c c}
				\hline\hline
				
				Shape & Class & $L$ & $w$ & $\overline{L}_h$ & $A$ & $P_x$ & $P_y$ \\
				$\ldots$    &  $\ldots$    & $m$ &  $m$ &  $m$ & $m^2$ & $m$ & $m$ \\
				1 & Barchan & 246 & 568 & 221 & 142577 & 1245 & 791 \\
				2 & Barchan & 219 & 578 & 257 & 140377 & 408 & 239 \\
				3 & Barchan & 216 & 510 & 269 & 115095 & 1992 & 1665 \\
				4 & Barchan & 145 & 309 & 231 & 55994 & 858 & 1294  \\
				\hline\hline
		\end{tabular}
	\end{center}
\end{table}

Having confirmed that the instance segmentation based on YOLOv8 successfully identify, classify, outline, and track each dune appearing on images from experiments, we trained the same CNN using satellite images of eolian dunes on the Martian and Earth's surfaces following a procedure described in the Methods section. After training a given set of images, we used the trained CNN to identify dunes in other images. For instance, Fig. \ref{fig:satellite} shows a HiRISE image \cite{Supplemental_HIRISE} of a field of individual barchans on the surface of Mars (23.190$^\circ$ latitude, 339.585 $^\circ$ longitude), where we can observe single barchans migrating over irregular terrains containing craters, while Fig. \ref{fig:satellite2}
shows a medium-resolution image (4 m per px) from CTX \cite{Supplemental_CTX} of a field of barchans undergoing complex interactions on the surface of Mars (-41.488$^\circ$ latitude, 44.589 $^\circ$ longitude). As for the experiments, the trained CNN is able to correctly identify, classify and outline dunes in satellite images with confidence scores of the order of 0.9 in the case of single barchans, and with lower accuracy in the case of interacting barchans, using both high- and medium-definition images. We note that the detection is not perfect in Fig. \ref{fig:satellite2}, with some dunes undergoing complex interactions not being detected while others are (all barchans are detected). However, we have shown from our experimental dunes (for which we have large datasets) that it is possible to have accurate detections of interacting barchans. For satellite images, datasets of interacting barchans are relatively small (time sequences that show the complete outcome of each interaction being absent) since a single barchan-barchan interaction on Earth takes decades to finish completely (on Mars it can take millenniums). This, added to the fact that the satellite images used are of lower quality than those from our experiments (in terms of spatial resolution and contrast with the background), decreases the detection accuracy of interacting dunes in Fig. \ref{fig:satellite2}.

Figure \ref{fig:satellite3} shows an image of a barchan field on Earth (24.836$^\circ$ latitude, 51.311$^\circ$ longitude, in Qatar), in which the contrast of colors between the dunes and background is poor. However, we observe that the CNN is able to detect and classify dunes with confidence scores of approximately 0.90 (the lower confidence score is 0.88, corresponding to a highly asymmetric barchan that is probably shedding a small dune through one of its horns), and to successfully outline them. Based on the outlines generated by the trained CNN, we measured the main features of all barchans \corr{identified in Figs.} \ref{fig:satellite}, \ref{fig:satellite2} and \ref{fig:satellite3}, which are listed in Tabs. \ref{tab:satellite1}, \ref{tab:satellite2} and \ref{tab:satellite3}, respectively.

We used the same procedure for other satellite images of Mars and Earth, as well as aerial pictures of eolian dunes, with different backgrounds (terrains), colors, resolutions, and point-of-view (view in perspective), and the results were as good as those shown in Figs. \ref{fig:satellite} and \ref{fig:satellite3}. In particular, we processed images with medium to low resolutions (15 m per px to 30 m per px) in which groups of barchans were undergoing complex interactions, and the trained CNN was able to correctly identify each dune (examples of instance segmentation of other satellite images are available in the Supplementary Information). The mean average precision $mAP$ (definition available in section Methods) reached in our CNN training was around 0.90 for the satellite images and 0.99 for the images from experiments (graphics of the evolution of $mAP$ along the epochs are available in the Supplementary Information). Finally, we carried out tracking (with detection and outline) in medium- to low-quality satellite images of an eight-year sequence of barchans undergoing dune-dune interactions in the Sahara desert, which we show in the Supplementary Information. Although the sequence contains only a small portion of barchan-barchan interactions (given the large timescales involved), the results are good, showing a great potential of the CNN for field tracking and measurements.





 

\clearpage

\section*{Discussion}

We carried out training of a single-stage object detection model YOLOv8 (YOLO version 8), together with scrips written in the course of this work to handle data and measure barchan dimensions, for image segmentation and tracking of barchan dunes. Different from previous works, we used a large database of time-resolved images of barchans of different sizes, colors, grain types, and format (camera type), consisting of mono or bidisperse grains (with more than one color), and undergoing different types of interaction \cite{Assis}. A small part of these images were used for training, and we afterward employed the trained CNN for processing images that were new to the CNN. With that, we could, besides the identification, classification, outline, and tracking of dunes, measure the time evolution of morphological quantities and compare them with the results from our non-machine learning detection code. For the experiments, the confidence scores were over 0.9, even when dunes of different color underwent different types of interaction. Therefore, for the first time, we showed the ability of a trained CNN to correctly identify, classify, outline and track dunes that undergo complex interactions with each other in a dune field, while previous works relied only on static satellite images for identifying single barchans.

We used the same procedure with satellite or aerial images of barchan fields on Mars and on Earth, with different image types, colors and perspectives, and in those cases the trained CNN identified, classified and outlined dunes with confidence scores above 70\%. However, in this case we did not systematically track dunes because of the small number of sequential images: good-quality images on Earth date back 30 years ago only, while dunes take a decade to displace a considerable distance, and on Mars the timescale is much higher (centuries or even millenniums). In the particular case of barchan-barchan interactions, there is no image sequence from satellites showing the entire process (only part of it) since time scales are much higher than those for subaqueous barchans (it would need a century or more of satellite images \corr{from Earth, and even more from Mars}, to finish the typical barchan-barchan interactions \cite{Assis}). We nevertheless carried out tracking (with detection and outline) in medium- to low-quality satellite images of an eight-year sequence of barchans undergoing dune-dune interactions in the Sahara desert, and the result was good (the results are available in the Supplementary Information), showing the potential of the technique for field measurements.

Although considerable improvements have been achieved in the automatic detection of barchans undergoing complex interactions, the trained CNN has still an important limitation: when processing images where barchans interact over a terrain (background) that has poor contrast with respect to the dunes, some dunes are not detected, and those detected have lower accuracy (such as happens in Fig. \ref{fig:satellite2}). When the contrast with the background is good and image resolution is not poor, the barchans are correctly identified with high accuracy (as in Figs. \ref{fig:detection} and \ref{fig:detection2}). However, as can be seen in Figs. \ref{fig:satellite} to \ref{fig:satellite3} and in those in the Supplementary Information, the great majority of barchans is correctly identified, classified, and outlined.

The success in identifying, classifying, outlining, and tracking barchans undergoing complex interactions by using CNN can be employed for dune monitoring, which engenders positive impacts in human activities. For example, it can be used for monitoring the growth and migration of dunes that are burying (or on the verge of burying) human constructions, such as in Florianopolis (Brazil) and Silver Lake (USA) \cite{Gaertner, WoodTV}, or for detecting complex barchan interactions on Mars. Besides, it can be explored further: the CNN can be trained on the history of certain barchan-barchan patterns (based on experimental data such as \corr{Refs.} \cite{Assis, Supplemental_binary_grl}), and the trained CNN can be afterward applied, for example, for deducing the past history of barchan fields on Mars (based on satellite images openly available). It can be also used for predicting the future of those barchan fields. If (or when) carried out, this would represent a considerable step for understanding the ancient past of Mars, for comprehending the undergoing climate change on Earth \cite{Baas}, and for predicting the far future of our planet. Our results represent, therefore, an important step in that direction.

\clearpage

\section*{Methods}

\subsection*{Experimental setup}

\begin{figure}[ht]
	\centering
	\includegraphics[width=0.9\linewidth]{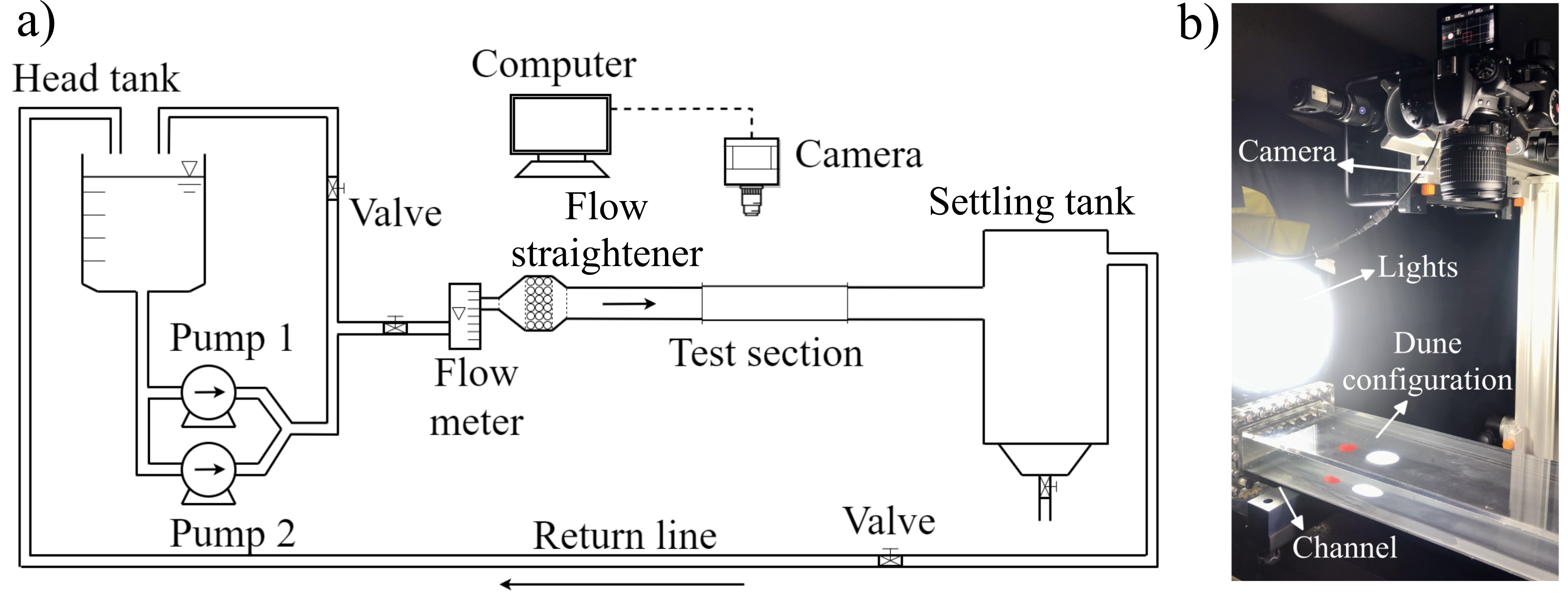}
	\caption{(a) Layout of the experimental setup. (b) Photograph of the test section.}
	\label{fig:experimental_setup}
\end{figure}

The CNN was trained with images from controlled experiments. The experimental setup consisted of a water tank, two centrifugal pumps, a flow straightener, a 5-m-long closed-conduit channel, a settling tank, and a return line, and we imposed a pressure-driven water flow in closed loop following the aforementioned order. The channel had a rectangular cross section 160 mm wide by 2$\delta$ = 50 mm high and was made of transparent material. It consisted of a 3-m-long entrance section (corresponding to 40 hydraulic diameters), a 1-m-long test section, and a 1-m-long section connecting the test section to the channel exit. With the channel completely filled with water in still conditions, controlled grains were poured inside, forming one or more conical heaps. Afterward, we imposed a specified water flow which deformed each conical pile into a barchan dune and, in the case of multiple piles, the barchan dunes interacted with each other.  We used tap water at temperatures within 22 and 30 $^{\circ}$C and different populations of grains (sometimes mixed): round glass beads ($\rho_s$ = 2500 kg/m$^3$) with $0.15$ mm $\leq\,d\,\leq$ $0.25$ mm and $0.40$ mm $\leq\,d\,\leq$ $0.60$ mm, angular glass beads with $0.21$ mm $\leq\,d\,\leq$ $0.30$ mm, and zirconium beads ($\rho_s$ = 4100 kg/m$^3$) with $0.40$ mm $\leq\,d\,\leq$ $0.60$, where $\rho_s$ and $d$ are, respectively, the density and diameter of grains. We used grains of different colors in order to track them during barchan-barchan interactions. A layout and a photograph of the experimental setup are shown in Fig. \ref{fig:experimental_setup}, and are also available in Assis and Franklin \cite{Assis, Assis2}.

Top view images of the dunes were acquired with either a high-speed or a conventional camera mounted on a traveling system and placed above the channel. The high-speed camera was of complementary metal-oxide-semiconductor (CMOS) type with maximum resolution of 2560 px $\times$ 1600 px at 800 Hz, and the conventional camera, also of CMOS type, had a maximum resolution of 1920 px $\times$ 1080 px at 60 Hz. Both the camera and traveling system were controlled by a computer, and we varied the field of view and the ROI (region of interest) in accordance with the number of dunes in the test and their velocity and those of grains. We mounted lenses of $60$ mm focal distance and F2.8 maximum aperture on the cameras and made use of LED (light-emitting diode) lamps branched to a continuous-current source to provide the necessary light while avoiding beating with the frequencies of cameras. More details about the experimental setup can be found in Refs. \cite{Alvarez, Alvarez3,  Alvarez4, Alvarez6, Assis, Assis2, Assis3} Datasets with the images and results of the experiments are available in open repositories \cite{Supplemental_binary_grl, Supplemental_binary_jgr, Supplemental_binary_bidisp, Supplemental_exp_esteban}.

\subsection*{Object detection model using convolutional neural network}

\begin{figure}[ht]
	\centering
	\includegraphics[width=0.9\linewidth]{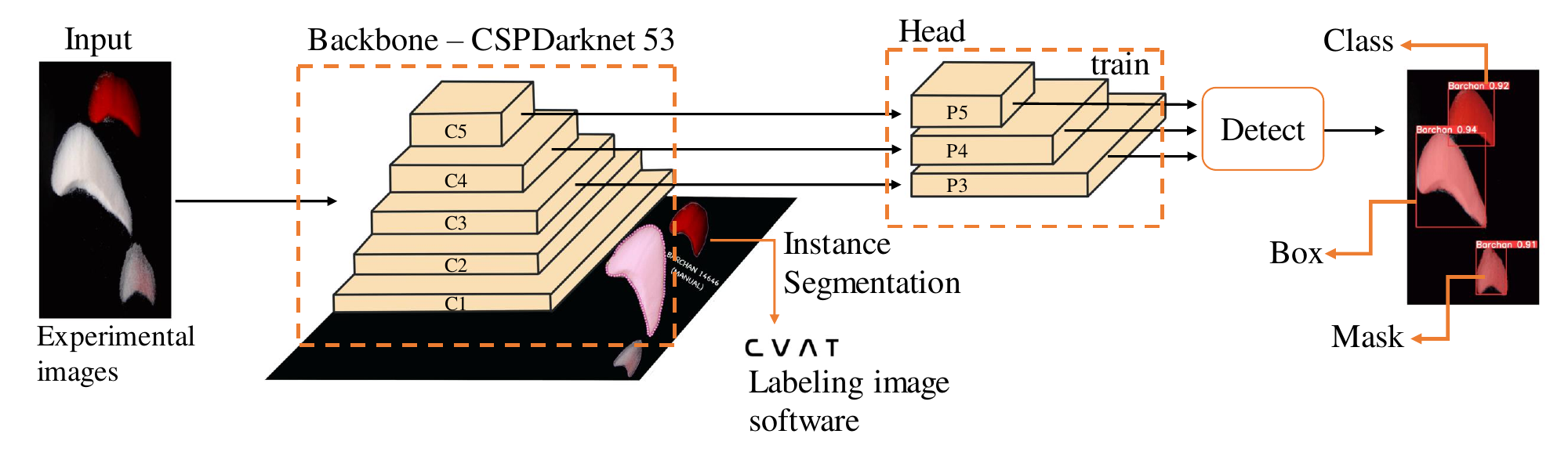}
	\caption{Flowchart showing a simplified architecture of YOLOv8. The C's represent the convolution layers (used in the backbone to extract relevant features from images, such as edges and textures) and the P's represent the pyramid of characteristics (used in the head to capture information at different scales and resolution).}
	\label{fig:yolo}
\end{figure}

We used the python library YOLOv8 (YOLO - You Only Look Once version 8) for carrying out instance segmentation of dunes (objects), in order to identify, classify, outline, and track each object along images of a given time sequence \cite{Yue}. YOLOv8 is \corr{a single-stage} object detection model based on CNN, whose architecture consists of backbone, neck and head, and it is known for fast generating masks while computing in parallel their coefficients \cite{Aboah}. The backbone (here the CSPDarknet-53) contains the CNN and generates feature maps at different levels of detail, which are passed to the neck. The neck then \corr{processes} the feature maps and builds feature maps for prediction, which are passed to the head. Finally, the head predicts the classes of objects, their bounding boxes, \corr{and their masks}, which can be directly used to outline objects. Figure \ref{fig:yolo} shows a simplified architecture of YOLOv8.

For the automatic dune detection, the following steps were taken sequentially. First, a large database of experimental data was constructed, taking into account binary interactions from previous work. Next, 7455 images were labeled using the CVAT platform (https://www.cvat.ai/) to annotate the images (this process was carried out manually), where we decided when dunes merged or a new dune was ejected based on the continuity of areas covered with grains (examples of object labeling with CVAT are available in the Supplementary Information). A validation and training database was then created using the labeled images, in which we used 876 images for validation, and we trained 300 epochs with a batch size equal to 1065. For the training, we made use of the pre-trained model yolov8n.pt, and trained two specific layers: one to detect barchan dunes (Barchan) and the other to detect non-barchan objects (Not a barchan). Finally, a Python code was developed to run the trained model, detect, classify and outline dunes, and analyze their morphology. The training was carried out in a GPU nvidia RTX 2070 using CUDA 12 and cuDNN 8 (CUDA Deep Neural Network library version 8), and we have not used data augmentation. However, images had different resolution, sharpness, and orientation.

The average accuracy of the trained CNN, for a given image dataset, is usually measured by the mean average precision $mAP$,

\begin{equation}
	mAP= \frac{1}{T}\Sigma_{i=1}^T AP_i \,\,,
\end{equation}

\noindent where $T$ is the number of categories (segmented trees) and $AP$ is the average precision of segmentation \cite{Yue, Soylu}. \corr{For a given category $i$, the average precision is given by the integral of the precision $P$ as a function of the recall $R$,}

\begin{equation}
	AP_i= \int_{0}^{1} PdR \,\,,
\end{equation}

\noindent where

\begin{equation}
	P = \frac{TP}{TP+FP} \qquad , \qquad R = \frac{TP}{TP+FN} \,\,,
\end{equation}

\noindent \corr{$TP$ being the true positives, $FP$ the false positives, and $FN$ the false negatives.} Finally, the intersection over Union $IoU$ is a measure of how much the detection box overlaps a box containing the real object (ground truth) \cite{Soylu}:

\begin{equation}
	IoU = \frac{intersection \,\, area}{union \,\, area} \,\,.
\end{equation}

The estimated accuracy $C$ plotted in Figs.\ref{fig:detection} and  \ref{fig:detection2}--\ref{fig:satellite2} for each detection box is usually called \textit{confidence score}, and corresponds to the precision $P$ multiplied by the $IoU$ and by the conditional class probability $P_c$:

\begin{equation}
	C = P_c \cdot P \cdot IoU \,\,,
\end{equation}

\noindent where $P_c$ indicates if a given class is present in the box. The trained CNN is available in an open repository \cite{Supplemental_cnn_exp}.

\subsection*{Satellite images}

We used a combination of satellite imagery platforms to collect images of the surfaces of Earth and Mars, which we used to train and apply the single-stage object detection model for identifying, classifying, and outlining dunes in satellite images. On Mars, high-resolution images were obtained from the HiRISE project \cite{Supplemental_HIRISE}, and medium- and low-resolution images from the Global CTX mosaic \cite{Supplemental_CTX}. For high resolution images, we made use of the HiView code \cite{Supplemental_HIRISE} for converting the pixel scale to a real physical unit (m). For terrestrial dunes, images ranging from low to high resolution were obtained from the Google Earth Pro and Copernicus \cite{Supplemental_Copernicus} platforms.

We trained the CNN following the same procedure as for the experiments (examples of object labeling with CVAT are available in the Supplementary Information). In this case, 12395 images were labeled and trained with 300 epochs, and 2376 images were used for validation. The trained CNN is available in an open repository \cite{Supplemental_cnn_sat}.


\section*{Data availability}

Data supporting our findings are available in open repositories \cite{Supplemental_binary_grl, Supplemental_binary_jgr, Supplemental_binary_bidisp, Supplemental_HIRISE, Supplemental_CTX, Supplemental_Copernicus, Supplemental_exp_esteban, Supplemental_cnn_exp, Supplemental_cnn_sat}.

\bibliography{references}

\begin{thebibliography}{10}
\urlstyle{rm}
\expandafter\ifx\csname url\endcsname\relax
  \def\url#1{\texttt{#1}}\fi
\expandafter\ifx\csname urlprefix\endcsname\relax\def\urlprefix{URL }\fi
\expandafter\ifx\csname doiprefix\endcsname\relax\def\doiprefix{DOI: }\fi
\providecommand{\bibinfo}[2]{#2}
\providecommand{\eprint}[2][]{\url{#2}}

\bibitem{Bagnold_1}
\bibinfo{author}{Bagnold, R.~A.}
\newblock \emph{\bibinfo{title}{The Physics of Blown Sand and Desert Dunes}}
  (\bibinfo{publisher}{Chapman and Hall}, \bibinfo{address}{London},
  \bibinfo{year}{1941}).

\bibitem{Hersen_1}
\bibinfo{author}{Hersen, P.}, \bibinfo{author}{Douady, S.} \&
  \bibinfo{author}{Andreotti, B.}
\newblock \bibinfo{journal}{\bibinfo{title}{Relevant length scale of barchan
  dunes}}.
\newblock {\emph{\JournalTitle{Phys. Rev. Lett.}}}
  \textbf{\bibinfo{volume}{89}}, \bibinfo{pages}{264301},
  \doiprefix\url{10.1103/PhysRevLett.89.264301} (\bibinfo{year}{2002}).

\bibitem{Rubanenko2}
\bibinfo{author}{Rubanenko, L.}, \bibinfo{author}{Lapôtre, M. G.~A.},
  \bibinfo{author}{Ewing, R.~C.}, \bibinfo{author}{Fenton, L.~K.} \&
  \bibinfo{author}{Gunn, A.}
\newblock \bibinfo{journal}{\bibinfo{title}{A distinct ripple-formation regime
  on {M}ars revealed by the morphometrics of barchan dunes}}.
\newblock {\emph{\JournalTitle{Nat. Commun.}}} \textbf{\bibinfo{volume}{13}},
  \doiprefix\url{10.1038/s41467-022-34974-3} (\bibinfo{year}{2022}).

\bibitem{Chojnacki2}
\bibinfo{author}{Chojnacki, M.}, \bibinfo{author}{Banks, M.~E.},
  \bibinfo{author}{Fenton, L.~K.} \& \bibinfo{author}{Urso, A.~C.}
\newblock \bibinfo{journal}{\bibinfo{title}{{Boundary condition controls on the
  high-sand-flux regions of {M}ars}}}.
\newblock {\emph{\JournalTitle{Geology}}} \textbf{\bibinfo{volume}{47}},
  \bibinfo{pages}{427--430}, \doiprefix\url{10.1130/G45793.1}
  (\bibinfo{year}{2019}).

\bibitem{Claudin_Andreotti}
\bibinfo{author}{Claudin, P.} \& \bibinfo{author}{Andreotti, B.}
\newblock \bibinfo{journal}{\bibinfo{title}{A scaling law for aeolian dunes on
  {M}ars, {V}enus, {E}arth, and for subaqueous ripples}}.
\newblock {\emph{\JournalTitle{Earth Plan. Sci. Lett.}}}
  \textbf{\bibinfo{volume}{252}}, \bibinfo{pages}{20--44}
  (\bibinfo{year}{2006}).

\bibitem{Hersen_2}
\bibinfo{author}{Hersen, P.} \emph{et~al.}
\newblock \bibinfo{journal}{\bibinfo{title}{Corridors of barchan dunes:
  Stability and size selection}}.
\newblock {\emph{\JournalTitle{Phys. Rev. E}}} \textbf{\bibinfo{volume}{69}},
  \bibinfo{pages}{011304}, \doiprefix\url{10.1103/PhysRevE.69.011304}
  (\bibinfo{year}{2004}).

\bibitem{Hersen_5}
\bibinfo{author}{Hersen, P.} \& \bibinfo{author}{Douady, S.}
\newblock \bibinfo{journal}{\bibinfo{title}{Collision of barchan dunes as a
  mechanism of size regulation}}.
\newblock {\emph{\JournalTitle{Geophys. Res. Lett.}}}
  \textbf{\bibinfo{volume}{32}} (\bibinfo{year}{2005}).

\bibitem{Kocurek}
\bibinfo{author}{Kocurek, G.}, \bibinfo{author}{Ewing, R.~C.} \&
  \bibinfo{author}{Mohrig, D.}
\newblock \bibinfo{journal}{\bibinfo{title}{How do bedform patterns arise? new
  views on the role of bedform interactions within a set of boundary
  conditions}}.
\newblock {\emph{\JournalTitle{Earth Surf. Process. Landforms}}}
  \textbf{\bibinfo{volume}{35}}, \bibinfo{pages}{51--63}
  (\bibinfo{year}{2010}).

\bibitem{Genois}
\bibinfo{author}{G\'enois, M.}, \bibinfo{author}{Hersen, P.},
  \bibinfo{author}{du~Pont, S.} \& \bibinfo{author}{Gr\'egoire, G.}
\newblock \bibinfo{journal}{\bibinfo{title}{Spatial structuring and size
  selection as collective behaviours in an agent-based model for barchan
  fields}}.
\newblock {\emph{\JournalTitle{Eur. Phys. J. B}}} \textbf{\bibinfo{volume}{86}}
  (\bibinfo{year}{2013}).

\bibitem{Genois2}
\bibinfo{author}{G\'enois, M.}, \bibinfo{author}{du~Pont, S.~C.},
  \bibinfo{author}{Hersen, P.} \& \bibinfo{author}{Gr\'egoire, G.}
\newblock \bibinfo{journal}{\bibinfo{title}{An agent-based model of dune
  interactions produces the emergence of patterns in deserts}}.
\newblock {\emph{\JournalTitle{Geophys. Res. Lett.}}}
  \textbf{\bibinfo{volume}{40}}, \bibinfo{pages}{3909--3914}
  (\bibinfo{year}{2013}).

\bibitem{Assis}
\bibinfo{author}{Assis, W.~R.} \& \bibinfo{author}{Franklin, E.~M.}
\newblock \bibinfo{journal}{\bibinfo{title}{A comprehensive picture for binary
  interactions of subaqueous barchans}}.
\newblock {\emph{\JournalTitle{Geophys. Res. Lett.}}}
  \textbf{\bibinfo{volume}{47}}, \bibinfo{pages}{e2020GL089464},
  \doiprefix\url{https://doi.org/10.1029/2020GL089464} (\bibinfo{year}{2020}).

\bibitem{Assis2}
\bibinfo{author}{Assis, W.~R.} \& \bibinfo{author}{Franklin, E.~M.}
\newblock \bibinfo{journal}{\bibinfo{title}{Morphodynamics of barchan-barchan
  interactions investigated at the grain scale}}.
\newblock {\emph{\JournalTitle{J. Geophys. Res.: Earth Surf.}}}
  \textbf{\bibinfo{volume}{126}}, \bibinfo{pages}{e2021JF006237},
  \doiprefix\url{https://doi.org/10.1029/2021JF006237} (\bibinfo{year}{2021}).

\bibitem{Assis3}
\bibinfo{author}{Assis, W.~R.}, \bibinfo{author}{Cúñez, F.~D.} \&
  \bibinfo{author}{Franklin, E.~M.}
\newblock \bibinfo{journal}{\bibinfo{title}{Revealing the intricate dune-dune
  interactions of bidisperse barchans}}.
\newblock {\emph{\JournalTitle{J. Geophys. Res.: Earth Surf.}}}
  \textbf{\bibinfo{volume}{127}}, \bibinfo{pages}{e2021JF006588},
  \doiprefix\url{https://doi.org/10.1029/2021JF006588} (\bibinfo{year}{2022}).

\bibitem{Parteli5}
\bibinfo{author}{Parteli, E. J.~R.}, \bibinfo{author}{Durán, O.},
  \bibinfo{author}{Tsoar, H.}, \bibinfo{author}{Schwämmle, V.} \&
  \bibinfo{author}{Herrmann, H.~J.}
\newblock \bibinfo{journal}{\bibinfo{title}{Dune formation under bimodal
  winds}}.
\newblock {\emph{\JournalTitle{Proc. Natl. Acad. Sci. U.S.A.}}}
  \textbf{\bibinfo{volume}{106}}, \bibinfo{pages}{22085--22089},
  \doiprefix\url{10.1073/pnas.0808646106} (\bibinfo{year}{2009}).
\newblock \eprint{https://www.pnas.org/doi/pdf/10.1073/pnas.0808646106}.

\bibitem{Courrech2}
\bibinfo{author}{Courrech~du Pont, S.}, \bibinfo{author}{Narteau, C.} \&
  \bibinfo{author}{Gao, X.}
\newblock \bibinfo{journal}{\bibinfo{title}{Two modes for dune orientation}}.
\newblock {\emph{\JournalTitle{Geology}}} \textbf{\bibinfo{volume}{42}},
  \bibinfo{pages}{743--746}, \doiprefix\url{10.1130/G35657.1}
  (\bibinfo{year}{2014}).
\newblock
  \eprint{https://pubs.geoscienceworld.org/gsa/geology/article-pdf/42/9/743/3546522/743.pdf}.

\bibitem{Gadal}
\bibinfo{author}{Gadal, C.}, \bibinfo{author}{Narteau, C.},
  \bibinfo{author}{du~Pont, S.~C.}, \bibinfo{author}{Rozier, O.} \&
  \bibinfo{author}{Claudin, P.}
\newblock \bibinfo{journal}{\bibinfo{title}{Incipient bedforms in a
  bidirectional wind regime}}.
\newblock {\emph{\JournalTitle{J. Fluid Mech.}}}
  \textbf{\bibinfo{volume}{862}}, \bibinfo{pages}{490--516}
  (\bibinfo{year}{2019}).

\bibitem{Bourke2}
\bibinfo{author}{Bourke, M.~C.} \& \bibinfo{author}{Goudie, A.~S.}
\newblock \bibinfo{journal}{\bibinfo{title}{Varieties of barchan form in the
  namib desert and on mars}}.
\newblock {\emph{\JournalTitle{Aeolian Research}}}
  \textbf{\bibinfo{volume}{1}}, \bibinfo{pages}{45--54},
  \doiprefix\url{https://doi.org/10.1016/j.aeolia.2009.05.002}
  (\bibinfo{year}{2009}).

\bibitem{Silvestro}
\bibinfo{author}{Silvestro, S.}, \bibinfo{author}{Vaz, D.~A.},
  \bibinfo{author}{Fenton, L.~K.} \& \bibinfo{author}{Geissler, P.~E.}
\newblock \bibinfo{journal}{\bibinfo{title}{Active aeolian processes on mars: A
  regional study in arabia and meridiani terrae}}.
\newblock {\emph{\JournalTitle{Geophys. Res. Lett.}}}
  \textbf{\bibinfo{volume}{38}},
  \doiprefix\url{https://doi.org/10.1029/2011GL048955} (\bibinfo{year}{2011}).

\bibitem{Fenton}
\bibinfo{author}{Fenton, L.~K.}
\newblock \bibinfo{journal}{\bibinfo{title}{Updating the global inventory of
  dune fields on mars and identification of many small dune fields}}.
\newblock {\emph{\JournalTitle{Icarus}}} \textbf{\bibinfo{volume}{352}},
  \bibinfo{pages}{114018},
  \doiprefix\url{https://doi.org/10.1016/j.icarus.2020.114018}
  (\bibinfo{year}{2020}).

\bibitem{Tsoar}
\bibinfo{author}{Tsoar, H.}, \bibinfo{author}{Greeley, R.} \&
  \bibinfo{author}{Peterfreund, A.~R.}
\newblock \bibinfo{journal}{\bibinfo{title}{{MARS}: {T}he north polar sand sea
  and related wind patterns}}.
\newblock {\emph{\JournalTitle{J. Geophys. Res.}}}
  \textbf{\bibinfo{volume}{84}}, \bibinfo{pages}{8167--8180},
  \doiprefix\url{https://doi.org/10.1029/JB084iB14p08167}
  (\bibinfo{year}{1979}).

\bibitem{Tsoar2}
\bibinfo{author}{Tsoar, H.} \& \bibinfo{author}{Parteli, E. J.~R.}
\newblock \bibinfo{journal}{\bibinfo{title}{Bidirectional winds, barchan dune
  asymmetry and formation of seif dunes from barchans: a discussion}}.
\newblock {\emph{\JournalTitle{Environ. Earth Sci.}}}
  \textbf{\bibinfo{volume}{75}},
  \doiprefix\url{https://doi.org/10.1007/s12665-016-6040-4}
  (\bibinfo{year}{2016}).

\bibitem{Zhang_2}
\bibinfo{author}{Zhang, Z.}, \bibinfo{author}{Dong, Z.}, \bibinfo{author}{Hu,
  G.} \& \bibinfo{author}{Parteli, E. J.~R.}
\newblock \bibinfo{journal}{\bibinfo{title}{Migration and morphology of
  asymmetric barchans in the central hexi corridor of northwest china}}.
\newblock {\emph{\JournalTitle{Geosciences}}} \textbf{\bibinfo{volume}{8}},
  \doiprefix\url{10.3390/geosciences8060204} (\bibinfo{year}{2018}).

\bibitem{Azzaoui}
\bibinfo{author}{Azzaoui, M.~A.}, \bibinfo{author}{Adnani, M.},
  \bibinfo{author}{El~Belrhiti, H.}, \bibinfo{author}{Chaouki, I.~E.} \&
  \bibinfo{author}{Masmoudi, C.}
\newblock \bibinfo{journal}{\bibinfo{title}{Detection of barchan dunes in high
  resolution satellite images}}.
\newblock {\emph{\JournalTitle{The International Archives of the
  Photogrammetry, Remote Sensing and Spatial Information Sciences}}}
  \textbf{\bibinfo{volume}{XLI-B7}}, \bibinfo{pages}{153--160},
  \doiprefix\url{10.5194/isprs-archives-XLI-B7-153-2016}
  (\bibinfo{year}{2016}).

\bibitem{Carrera}
\bibinfo{author}{Carrera, D.}, \bibinfo{author}{Bandeira, L.},
  \bibinfo{author}{Santana, R.} \& \bibinfo{author}{Lozano, J.~A.}
\newblock \bibinfo{journal}{\bibinfo{title}{Detection of sand dunes on {M}ars
  using a regular vine-based classification approach}}.
\newblock {\emph{\JournalTitle{Knowledge-Based Systems}}}
  \textbf{\bibinfo{volume}{163}}, \bibinfo{pages}{858--874},
  \doiprefix\url{https://doi.org/10.1016/j.knosys.2018.10.011}
  (\bibinfo{year}{2019}).

\bibitem{Rubanenko}
\bibinfo{author}{Rubanenko, L.}, \bibinfo{author}{Pérez-López, S.},
  \bibinfo{author}{Schull, J.} \& \bibinfo{author}{Lapôtre, M. G.~A.}
\newblock \bibinfo{journal}{\bibinfo{title}{Automatic detection and
  segmentation of barchan dunes on mars and earth using a convolutional neural
  network}}.
\newblock {\emph{\JournalTitle{IEEE J. Sel. Top. Appl.}}}
  \textbf{\bibinfo{volume}{14}}, \bibinfo{pages}{9364--9371},
  \doiprefix\url{10.1109/JSTARS.2021.3109900} (\bibinfo{year}{2021}).

\bibitem{kowalczyk_sup}
\bibinfo{author}{Kowalczyk, A.}
\newblock \emph{\bibinfo{title}{Support vector machines succinctly}}
  (\bibinfo{publisher}{Syncfusion Inc}, \bibinfo{year}{2017}).

\bibitem{He}
\bibinfo{author}{He, K.}, \bibinfo{author}{Gkioxari, G.},
  \bibinfo{author}{Dollár, P.} \& \bibinfo{author}{Girshick, R.}
\newblock \bibinfo{title}{Mask r-cnn}.
\newblock In \emph{\bibinfo{booktitle}{2017 IEEE International Conference on
  Computer Vision (ICCV)}}, \bibinfo{pages}{2980--2988},
  \doiprefix\url{10.1109/ICCV.2017.322} (\bibinfo{year}{2017}).

\bibitem{Baas}
\bibinfo{author}{Baas, A. C.~W.} \& \bibinfo{author}{Delobel, L.~A.}
\newblock \bibinfo{journal}{\bibinfo{title}{Desert dunes transformed by
  end-of-century changes in wind climate}}.
\newblock {\emph{\JournalTitle{Nat. Clim. Chang.}}}
  \textbf{\bibinfo{volume}{12}}, \bibinfo{pages}{999--1006},
  \doiprefix\url{10.1038/s41558-022-01507-1} (\bibinfo{year}{2022}).

\bibitem{Assis4}
\bibinfo{author}{Assis, W.~R.}, \bibinfo{author}{Borges, D.~S.} \&
  \bibinfo{author}{Franklin, E.~M.}
\newblock \bibinfo{journal}{\bibinfo{title}{Barchan dunes cruising dune-size
  obstacles}}.
\newblock {\emph{\JournalTitle{Geophys. Res. Lett.}}}
  \textbf{\bibinfo{volume}{50}}, \bibinfo{pages}{e2023GL104125},
  \doiprefix\url{https://doi.org/10.1029/2023GL104125} (\bibinfo{year}{2023}).

\bibitem{WoodTV}
\bibinfo{author}{WOODTV8}.
\newblock \bibinfo{title}{Silver {L}ake {D}unes swallow up house}.
\newblock
  \bibinfo{howpublished}{\url{https://www.youtube.com/watch?v=ifxzsMA4IWY}}
  (\bibinfo{year}{2017}).

\bibitem{Supplemental_binary_grl}
\bibinfo{author}{Assis, W.~R.} \& \bibinfo{author}{Franklin, E.~M.}
\newblock \bibinfo{title}{Experimental data on binary interactions of
  subaqueous barchans}.
\newblock \bibinfo{howpublished}{\emph{Mendeley Data}
  \url{http://dx.doi.org/10.17632/jn3kt83hzh.3}} (\bibinfo{year}{2020}).

\bibitem{Supplemental_HIRISE}
\bibinfo{title}{High resolution imaging science experiment}.
\newblock \bibinfo{howpublished}{\emph{HiRISE Operations Center - University of
  Arizona} \url{https://www.actgate.com/}}.

\bibitem{Supplemental_CTX}
\bibinfo{title}{Mars reconnaissance orbiter (mro) context camera (ctx)
  dataset}.
\newblock \bibinfo{howpublished}{\emph{Applied Coherent Technology (ACT)
  Corporation} \url{https://www.uahirise.org/}}.

\bibitem{Gaertner}
\bibinfo{author}{Gaertner, E.}
\newblock \bibinfo{journal}{\bibinfo{title}{Lake {M}ichigan sand dune threatens
  to swallow another silver lake cottage}}.
\newblock {\emph{\JournalTitle{mlive}}}  (\bibinfo{year}{2017}).

\bibitem{Alvarez}
\bibinfo{author}{Alvarez, C.~A.} \& \bibinfo{author}{Franklin, E.~M.}
\newblock \bibinfo{journal}{\bibinfo{title}{Birth of a subaqueous barchan
  dune}}.
\newblock {\emph{\JournalTitle{Phys. Rev. E}}} \textbf{\bibinfo{volume}{96}},
  \bibinfo{pages}{062906}, \doiprefix\url{10.1103/PhysRevE.96.062906}
  (\bibinfo{year}{2017}).

\bibitem{Alvarez3}
\bibinfo{author}{Alvarez, C.~A.} \& \bibinfo{author}{Franklin, E.~M.}
\newblock \bibinfo{journal}{\bibinfo{title}{Role of transverse displacements in
  the formation of subaqueous barchan dunes}}.
\newblock {\emph{\JournalTitle{Phys. Rev. Lett.}}}
  \textbf{\bibinfo{volume}{121}}, \bibinfo{pages}{164503},
  \doiprefix\url{10.1103/PhysRevLett.121.164503} (\bibinfo{year}{2018}).

\bibitem{Alvarez4}
\bibinfo{author}{Alvarez, C.~A.} \& \bibinfo{author}{Franklin, E.~M.}
\newblock \bibinfo{journal}{\bibinfo{title}{Horns of subaqueous barchan dunes:
  A study at the grain scale}}.
\newblock {\emph{\JournalTitle{Phys. Rev. E}}} \textbf{\bibinfo{volume}{100}},
  \bibinfo{pages}{042904}, \doiprefix\url{10.1103/PhysRevE.100.042904}
  (\bibinfo{year}{2019}).

\bibitem{Alvarez6}
\bibinfo{author}{Alvarez, C.~A.}, \bibinfo{author}{C\'u\~nez, F.~D.} \&
  \bibinfo{author}{Franklin, E.~M.}
\newblock \bibinfo{journal}{\bibinfo{title}{Growth of barchan dunes of
  bidispersed granular mixtures}}.
\newblock {\emph{\JournalTitle{Phys. Fluids}}} \textbf{\bibinfo{volume}{33}},
  \bibinfo{pages}{051705} (\bibinfo{year}{2021}).

\bibitem{Supplemental_binary_jgr}
\bibinfo{author}{Assis, W.~R.} \& \bibinfo{author}{Franklin, E.~M.}
\newblock \bibinfo{title}{Experimental data on barchan-barchan interaction at
  the grain scale}.
\newblock \bibinfo{howpublished}{\emph{Mendeley Data}
  \url{http://doi.org/10.17632/f9p59sxm4f.1}} (\bibinfo{year}{2021}).

\bibitem{Supplemental_binary_bidisp}
\bibinfo{author}{Assis, W.~R.}, \bibinfo{author}{C\'u\~nez, F.} \&
  \bibinfo{author}{Franklin, E.~M.}
\newblock \bibinfo{title}{Experimental data on barchan-barchan interaction with
  bidisperse grains}.
\newblock \bibinfo{howpublished}{\emph{Mendeley Data}
  \url{http://doi.org/10.17632/sbjtzbzh9k.1}} (\bibinfo{year}{2021}).

\bibitem{Supplemental_exp_esteban}
\bibinfo{author}{C\'u\~nez, E.~A.} \& \bibinfo{author}{Franklin, E.~M.}
\newblock \bibinfo{title}{Experimental dataset on ``{D}etection and tracking of
  barchan dunes using artificial intelligence''}.
\newblock \bibinfo{howpublished}{\emph{Mendeley Data}
  \url{http://doi.org/10.17632/8wh3w3y899}} (\bibinfo{year}{2023}).

\bibitem{Yue}
\bibinfo{author}{Yue, X.} \emph{et~al.}
\newblock \bibinfo{journal}{\bibinfo{title}{Improved {YOLO}v8-seg network for
  instance segmentation of healthy and diseased tomato plants in the growth
  stage}}.
\newblock {\emph{\JournalTitle{Agriculture}}} \textbf{\bibinfo{volume}{13}},
  \doiprefix\url{10.3390/agriculture13081643} (\bibinfo{year}{2023}).

\bibitem{Aboah}
\bibinfo{author}{Aboah, A.}, \bibinfo{author}{Wang, B.},
  \bibinfo{author}{Bagci, U.} \& \bibinfo{author}{Adu-Gyamfi, Y.}
\newblock \bibinfo{title}{Real-time multi-class helmet violation detection
  using few-shot data sampling technique and {YOLOv8}}.
\newblock In \emph{\bibinfo{booktitle}{2023 {IEEE/CVF} Conference on Computer
  Vision and Pattern Recognition Workshops ({CVPRW})}},
  \bibinfo{pages}{5350--5358}, \doiprefix\url{10.1109/CVPRW59228.2023.00564}
  (\bibinfo{year}{2023}).

\bibitem{Soylu}
\bibinfo{author}{Soylu, E.} \& \bibinfo{author}{Soylu, T.}
\newblock \bibinfo{journal}{\bibinfo{title}{A performance comparison of
  {YOLOv8} models for traffic sign detection in the {R}obotaxi-full scale
  autonomous vehicle competition}}.
\newblock {\emph{\JournalTitle{Multimed Tools Appl.}}}
  \doiprefix\url{10.1007/s11042-023-16451-1} (\bibinfo{year}{2023}).

\bibitem{Supplemental_cnn_exp}
\bibinfo{author}{C\'u\~nez, E.~A.} \& \bibinfo{author}{Franklin, E.~M.}
\newblock \bibinfo{title}{{CNN} training of experimental images for
  ``{D}etection and tracking of barchan dunes using artificial intelligence''}.
\newblock \bibinfo{howpublished}{\emph{Mendeley Data}
  \url{http://doi.org/10.17632/brgxgtpz92}} (\bibinfo{year}{2023}).

\bibitem{Supplemental_Copernicus}
\bibinfo{title}{Copernicus eu}.
\newblock \bibinfo{howpublished}{\emph{Copernicus Browser}
  \url{https://browser.dataspace.copernicus.eu/}}.

\bibitem{Supplemental_cnn_sat}
\bibinfo{author}{C\'u\~nez, E.~A.} \& \bibinfo{author}{Franklin, E.~M.}
\newblock \bibinfo{title}{{CNN} training of satellite images for ``{D}etection
  and tracking barchan dunes using artificial intelligence''}.
\newblock \bibinfo{howpublished}{\emph{Mendeley Data}
  \url{http://doi.org/10.17632/v4yntwdnjk}} (\bibinfo{year}{2023}).

\end{thebibliography}



\section*{Acknowledgements}

The authors are grateful to FAPESP (Grant Nos. 2018/14981-7 and 2021/11470-4) and to CNPq (Grant No. 405512/2022-8) for the financial support provided. The authors thank Fernando David C\'un\~ez (University of Rochester) for the help with measuring objects in satellite images.

\section*{Author contributions statement}

E.A.C. wrote the numerical scripts, carried out the computations (and CNN training), carried out some of the experiments, and processed the data.  E.M.F conceived the work, analyzed the data, was responsible for the funding, and wrote the manuscript. Both authors reviewed the manuscript.


\section*{Additional information}

\textbf{Competing interests:} The authors declare no competing interests.





\end{document}